\documentclass[12pt]{article}
\usepackage{graphicx}
\newcommand{\al}{\alpha}
\newcommand{\rel}[1]{\frac{\textstyle 1}{\textstyle #1}}

\newcommand{\Li}{\mathop{\rm Li_2}\nolimits}

\title{Exponential Basis in Two-Sided Variational Estimates of
Energy for Three-Body Systems}
\author{A.G.Donchev, N.N.Kolesnikov, V.I.Tarasov}
\date{}
\begin{document}
\maketitle

\begin{abstract}
By the use of the variational method with exponential trial
functions the upper and lower bounds of energy are calculated for
a number of non-relativistic three-body Coulomb and nuclear
systems. The formulas for calculation of upper and lower bounds
for exponential basis are given, the lower bounds for great part
of systems were calculated for the first time. By comparison of
calculations for different bases the efficiency of exponential
trial functions and their universality in respect to masses of
particles and interaction are demonstrated. The advantage of
exponential basis manifests mostly evident for the systems with
comparable masses, though its use in one-center and
two-center problems is justified too. For effective
solution of two-center problem a carcass modification of the trial
function is proposed. The stability of various three-particle
Coulomb systems is analyzed.
\end{abstract}

\section{Introduction}
Among existing methods of calculation of non-relativistic  bounded
systems the variational method seems to be the most universal one as
it is applied equally well for the solution of atomic and nuclear
problems. It is essential  that the variational method allows to
find not only the upper ($E_U$) but also the lower ($E_L$)
estimates of the energy. As to potentiality of the method, there
are many examples of highly accurate calculations of
three- and more-particle systems \cite{Sch62}-\cite{Frl99}. For
instance in the three-body Coulomb problem the precision is
amounted up to a score of decimal places. Of course in real physical
systems the relativistic and other effects lead to corrections in
the energy already in 5--7-th decimal place and therefore, in
practical use a variational procedure which ensures a reasonable
accuracy with least computational efforts may be acceptable.

Historically the first variational expansion in a three-particle
problem was suggested by Hylleraas in perimetrical coordinates
in the form of exponent function multiplied by a polynomial with
integer nonnegative powers. Later negative and fractional powers
were added \cite{Sch62}, \cite{Thk94}, besides Frankowski and Pekeris
\cite{Pkr66} introduced logarithmic terms. In the next this basis
is referred as 'polynomial' one.

Another possibility is to use a purely exponential basis. It
ensures a good flexibility of the variational expansion due to the
presence of many scale nonlinear parameters. Whereas the Hylleraas
basis is practically oriented on solution of uni-center Coulomb
problems the exponential basis is good for systems with any masses
of particles and types of their interaction \cite{Fro92}. Besides, the
calculations with exponential basis are more simple and uniform
whereas for polynomial basis they became more and more complicated
as number of terms increases, especially if the logarithmic terms
are included.

Instead of exponents another non-polynomial functions, gaussians,
can be used. They are not less than exponents fitted for the
systems with any masses of particles, moreover they are applicable
to systems with arbitrary number of particles. For this basis all
the formulas needed for calculation of both the upper and lower
bounds are given in paper \cite{KTr82}, and different 3-, 4-, and
5-particle systems were calculated there. For the upper bound a
generalization is given in article \cite{Vrg98} for arbitrary orbital moments.  Nevertheless, our analysis have shown that at least for three-body variational calculations with not very high number of parameters the
precision for gaussian basis is lower than for exponential one.

Our principal goal was not the striving for improvement of
existing super-high precision calculations but the analysis of
efficiency of exponential and partly gaussian trial functions for
evaluation of the upper and lower variational bounds. For this
purpose the Coulomb and nuclear systems of particles with
different masses and types of interaction are considered and the
results of calculations are compared with those published in
literature.

To facilitate  such a comparison we will characterize the accuracy
of calculations of $E_U$ and $E_L$ by the values:
\begin{equation}
\label{deltas}
\delta_U=-\lg \left(\frac{E_0-E_U}{E_0}\right),\qquad
\delta_L=-\lg \left(\frac{E_L-E_0}{E_0}\right),
\end{equation}
which determine the number of correct decimal places of $E_U$ and
$E_L$ respectively, $E_0$ being the exact value of the energy.

The universality of exponential basis in respect to masses of
particles allows us to analyze the problem of stability of
different Coulomb three-particle systems.

\section{Method of calculation}
In three-particle problem it is convenient to use interparticle
distances as coordinates together with the Euler angles describing the
orientation of the triangle formed by the particles.
In the case of central interaction the wave function of
the ground state (and exited state with zero orbital momentum) depends
only on interparticle distances, therefore the function can
be written as:
\begin{equation}
\label{basis} \mid a\rangle = \exp\Big(-\sum\limits_{p=1}^3
\alpha_p^a R_p\Big),
\end{equation}
where $\alpha_p^a$ are the nonlinear variational parameters
specifying the scale of the basis function $\mid a\rangle$, $R_i$
is the distance between particles $j$ and $k$ where $\{i,j,k\}$ is
the triplet $\{1,2,3\}$ or its cyclic permutation. In the case of
gaussian basis $R_p$ in (\ref{basis}) is replaced by $R_p^2$.

It is convenient to use the notations:
$$
t_p \equiv \cos\beta_p                        , \hspace{1cm}
T_p \equiv \langle a\mid t_p     \mid b\rangle, \hspace{1cm}
G_p \equiv \langle a\mid R_p^{-1}\mid b\rangle, \hspace{1cm}
N   \equiv \langle a\mid              b\rangle,
$$
where $\beta_p$ is the angle at $p$-th particle in the triangle.
Then simple calculations result in the following formula for matrix
element of operator of kinetic energy, $T$, between states $\mid
a\rangle$ and $\mid b\rangle$:
$$
\langle a\mid T\mid b\rangle = \sum_{p=1}^3 s_p^a G_p -
                               \sum_{p=1}^3 d_p^a T_p - u^a N,
$$
where
$$
s_i^a=\al_i^a(\frac1{M}-\frac1{m_i}),
\hspace{1cm} u^a= \sum_{p=1}^{3} s_p^a \al_p^a, \hspace{1cm}
d_i^a=\frac{\al_1^a\al_2^a\al_3^a}{\al_i^am_i},
$$
$m_i$ being the mass of $i$-th particle and $1/M\equiv1/m_1+1/m_2+1/m_3$.

A calculation of matrix elements for the potential energy reduces
to calculation of the integrals similar to those for kinetic
energy. In particular, for the Coulomb interaction $V_p=1/R_p$:
$$
\langle a\mid V_p \mid b\rangle=G_p.
$$

A calculation of lower variational estimate requires additional
evaluations of matrix elements for operators $T^2, V^2$ and $VT$.
For this purpose it is convenient to introduce additional
notations:
$$
J_{pq} \equiv \langle a \mid R_p^{-1}R_q^{-1} \mid b\rangle,   \hspace{1cm}
W_{pq} \equiv \langle a \mid      t_pR_q^{-1} \mid b\rangle,   \hspace{1cm}
Q_{pq} \equiv \langle a \mid           t_pt_q \mid b\rangle.
$$

Then, the matrix elements of operators $T^2$ and $V_pT+TV_p$ are written as:
\begin{eqnarray*}
\langle a\mid T^2\mid b\rangle & = & u^a u^b N-
\sum_{p=1}^3 (s_p^a u^b + s_p^b u^a) G_p+
\sum_{p=1}^3 (d_p^a u^b + d_p^b u^a) T_p- \\& - &
\sum_{p,q=1}^3 (s_q^a d_p^b + s_q^b d_p^a) W_{pq}+
\sum_{p,q=1}^3 s_p^a s_q^b J_{pq}+
\sum_{p,q=1}^3 d_p^a d_q^b Q_{pq}.
\end{eqnarray*}
\begin{eqnarray*}
\langle a\mid V_pT+TV_p \mid b\rangle & = &
\sum_{q=1}^3 (s_q^a+s_q^b) \langle a\mid V_p/R_q \mid b\rangle -
\sum_{q=1}^3 (d_q^a+d_q^b) \langle a\mid V_p t_q \mid b\rangle - \\ & - &
             (u^a  + u^b ) \langle a\mid V_p     \mid b\rangle.
\end{eqnarray*}
In the particular case of Coulomb potential:
$$
\langle a\mid V_p/R_q \mid b\rangle = J_{pq}, \qquad
\langle a\mid V_p t_q \mid b\rangle = W_{qp}.
$$

The calculation of the matrix elements of $V^2$ is similar to
calculation of $\langle a\mid V\mid b\rangle$. In particular, for
the Coulomb interaction a simple formula takes place:
$$
\langle a\mid V_pV_q \mid b\rangle=J_{pq}.
$$

The trial function is written as a superposition of basis
functions (\ref{basis}):
\begin{equation}
\label{trial}
\psi=\sum_{a=1}^N C_a \mid a\rangle,
\end{equation}
where $C_a$ are linear parameters.

Evidently, the difficulties arise mostly at optimization of the
non-linear parameters. The possibilities of the deterministic
procedures are soon exhausted as the number of terms in expansion
(\ref{trial}) increases. Therefore, a specially designed procedure
of global stochastic searching was used. Briefly it is the
following: at each Monte-Carlo probe a random point is chosen in
$3N$-dimensional space of non-linear parameters according to
previously accepted distribution function. Then the coordinatewise
optimization is carried on, at first the stochastic one and then
the deterministic one. At this stage the best points are selected
for subsequent detailed optimization. Mentioned above distribution
function is found by a procedure similar to that described in
\cite{Don00}.

\section{Efficiency of calculations for various systems}
To understand better  what are the possibilities of exponential
basis and described above optimization procedure in calculations
of systems with different masses of particles and interactions a
number of Coulomb and nuclear systems were considered. Among them:
He atom, hydrogen ion H$^-$, positronium ion Ps$^-$ (e$^+e^-e^-$),
meso-systems $\alpha\mu^-e^-$, $p\mu^+e^-$, $pp\mu^-$,
$\mu^+\mu^+e^-$, $\mu^+e^-e^-$, two-center Coulomb systems
$ppe^-$, $dde^-$, $tte^-$, as well as nuclei $^3$H and
$_\Lambda^3$H.

The composition of the majority of considered Coulomb systems with
the particles of unit charge can be presented as $X^\pm Y^\mp
Y^\mp$, the identical particles being denoted as $Y$. The binding
energies decrease together with the values of masses but the accuracy
of calculation depends only on the ratio of masses. For these
systems the upper and lower bounds were calculated with $N=30$ in
expansion (\ref{trial}), corresponding values $\delta_U$ and
$\delta_L$ were plotted in Fig.~1 as the functions of mass ratio,
$\xi_{YX}=m_Y/m_X$. In calculations of the lower bound
the non-linear parameters were accepted to be equal to these found
for the upper bound.

As expected, the increase of ratio $\xi_{YX}$ leads to the
decrease of values $\delta_U$ and $\delta_L$ due to arising
difficulties in description of motion of heavy particles.
Nevertheless even at the approach to the two-center limit
($\xi_{YX} \gg 1$) the accuracy of calculations remains still
satisfactory. For comparison, in Fig.1 the results of most detailed
calculations with polynomial basis \cite{Kln90} are
presented too. It is seen that the accuracy of calculations with
polynomial basis \cite{Kln90} becomes bad for
$\xi_{YX}> 0.1$ in spite of large values of $N$. This comparison
shows that the exponential basis is applicable for a wider range
of values of $\xi_{YX}$ than the polynomial basis.

Note that in the case of Gaussian basis $\delta_U$ and $\delta_L$ decrease
even more slowly than for exponential basis (see Fig.1) though the latter provides generally higher precision.

The exponential basis can be used as well in the case of nuclear systems, even inconvenient for calculation (weakly bounded systems, short-range attractive potentials with strong repulsion at small distances between particles, that can be identical or not identical).
As a particularly 'inconvenient' system hypertritium,
$^3_\Lambda$H (consisting of $np\Lambda$), was chosen. For comparison a
more 'convenient' three-nucleon system, $^3$H, was considered. In
these calculations two types of model nuclear $NN$-potentials were
used, (i) purely attractive potential $NN-1$ and (ii) attractive potential with a soft core $NN-2$:
\begin{equation}
\label{YukawaPot}
 V_{NN}(r) = V_r(R_r/R) \exp (-R/R_r)-V_a(R_a/R) \exp (-R/R_a)
\end{equation}
and attractive $\Lambda N$-potential:
\begin{equation}
\label{LambdaNPot}
 V_{\Lambda N}(r) = -V_a \exp (-R/R_a),
\end{equation}
the parameters $V_{a,r}$ and $R_{a,r}$ are given in Appendix B.

The convergence of the upper and lower estimates for exponential
basis is illustrated in Table 1 and in Fig. 2 for various Coulomb and nuclear systems. As seen from Fig. 2 the dependence of
$\delta_U$ and $\delta_L$ on the $\lg N$ is close to the linear
one. In accordance with Fig. 1 the accuracy decreases as the
system approaches to the adiabatic limit, and in parallel the
convergence of variational estimates deteriorates (it is  characterized
by the slope of curves in Fig. 2). Note that the precision of
calculations for considered nuclear systems is generally similar
or even better than for Coulomb systems.

Besides, in Fig. 2 some results of calculations with the Gaussian
basis are shown (dotted lines). It is seen that the convergence of
the upper and lower bound is similar to that for exponential basis whereas
the accuracy is significantly lower.

\section{Comparison of results for different bases}
Comparison of efficiency of different variational expansions is
convenient to carry out on standard systems calculated by many
authors. Such systems are $^{\infty}$He and $^\infty$H$^-$ considered in
\cite{Sch62}-\cite{Pkr66},\cite{KTr82},\cite{Don00}-\cite{Frl98}.
In Fig. 3 the values of $\delta_U$ and $\delta_L$, are plotted for
these papers where the most detailed calculations of atom
$^\infty$He were carried out, the results of our calculations are
presented there too. Similarly, in Fig. 4 $\delta_U$ and
$\delta_L$ are presented  for hydrogen ion $^\infty$H$^-$.

It is necessary to emphasize that both cases are examples of
one-center systems. Therefore, this is a reason to expect that the
expansions especially designed for one-center problems will gain
the advantage. This is generally confirmed by our analysis. Up to
present the most accurate many-parameters calculations of
$^\infty$He were carried out using polynomial or
polynomial-logarithmic bases. As it is seen from Fig. 3 and Fig. 4
the convergence of the variational expansions for these bases is
generally better than for exponential or gaussian bases.

On the other hand, up to $\delta_U \approx 12$ the use of exponential basis is justified as it assures the same precision at lower number of terms (see Fig.3 and 4). Note that the over-high precision in non-relativistic calculations without taking into account relativistic and other corrections (that appears far before $\delta \approx 12$) have no physical meaning, though they are interesting from computational point of view.

As to the lower bound calculations they are rare in literature and
we estimate the number of $N$ up to which the calculations of
$E_L$ with exponential basis are justified (in the same sense as
for $E_U$) as $100-200$.

Another limiting case is the adiabatic one (i.e. a two-center system with two heavy
particles). In this case the use of polynomial basis leads to
unsatisfactory results, and the exponential basis is evidently
preferable (see Fig. 1). Moreover, the use of complex scale
parameters in exponential basis increases significantly the
accuracy of calculations \cite{FrS98}. The  most accurate
calculations of two-center systems were carried out in the
framework of the Born-Oppenheimer approach  \cite{BOp27} or its
modifications \cite{Pon81}, \cite{GDB98}. In particular, in paper
\cite{GDB98} the energy of the system $H_2^+$ ($ppe^-$) was calculated
with precision $\delta\approx 12$ but this is only some better
than that of the calculation of \cite{FrS98} with exponential
functions (note by the way that the number of basis terms in
\cite{FrS98} was less than in \cite{GDB98}).

A more effective modification of exponential basis in two-center
calculations is:
\begin{equation}
\label{carc}
|a\rangle=\exp(-\alpha_1^aR_1-\alpha_2^aR_2-\alpha_3^aR_3-\beta^aR_3^2),
\end{equation}
where $R_3$ is a distance between the heavy particles. Note that
the dependence of this function on $R_3$ can be presented as
$\exp(-\beta^a(R_3-R_3^a)^2)$, where $R_3^a$ is the new
variational parameter connected with $\alpha_3^a$.
Note that basis (\ref{carc}) is, in a certain sense, a particular case
of 'carcass' functions (constructed on the base of gaussians in
paper \cite{Zah83}), whose use together with gaussians might be
useful in nuclear physics for calculation with potentials changing
the sign.

For functions (\ref{carc}) all the integrals needed for calculations of the upper
variational bound are expressed in the closed form in terms of
conventional functions. For instance, the basic integral can be
calculated as:
\begin{equation}
G^{000} \equiv \int\limits_0^\infty e^{-x_3 R_3-\beta R_3^2} dR_3
\int\limits_0^\infty e^{-x_2 R_2} dR_2
\int\limits_{\mid R_2-R_3\mid}^{R_2+R_3} e^{-x_1 R_1} dR_1 =
-\frac1{\sqrt{\beta}}
\frac{F\Big(\frac{x_1+x_3}{2\sqrt{\beta}}\Big)-
F\Big(\frac{x_2+x_3}{2\sqrt{\beta}}\Big)}
{x_1^2-x_2^2},
\end{equation}
where $F(z)\equiv e^{z^2}\int_z^\infty e^{-t^2}dt$.

The calculations of the ground state of the system $ppe^-$ with
this modified basis lead to significantly better results than with
purely exponential or gaussian bases. In particular, in our
calculations it has been shown that even a single function
(\ref{carc}) provides a better precision than 50 exponents or
gaussians. Moreover, the basis (\ref{carc}) is more flexible than
the exponential basis with complex parameters used in
\cite{FrS98}. For instance, the result of calculations with $N=20$
for $ppe^-$ turns out to be better than that of paper \cite{FrS98}
with 200 complex exponents (1400 variational parameters) and
better than calculations with 300 functions for systems
$\mu^+\mu^+e^-$, $dde^-$ and $tte^-$.

In addition to the preceding discussion of two limiting cases
(one- and two-center problem) it is necessary to indicate that
there exists a large region of values of $\xi_{YX}$ between
$10^{-2}$ and $10^2$ where the exponential basis is beyond
compare. Note that this is the region where the great part of
known three-particle Coulomb systems is located. Thus, apart from
gaussians, the exponential basis seems to be the most universal one in
comparison with other approaches, applicable equally well to
Coulomb and nuclear three-particle systems.

\section{Stability of Coulomb Systems}
All considered above Coulomb systems except two ($\alpha e^-e^-$
and $\alpha\mu^-e^-$) had summary charge $\pm1$ and consisted of
three single-charged particles from which two are identical. All
systems of such type are stable in respect to separation of one of
the particles. However this is not the case for other type of
three particle Coulomb systems. For analysis of
stability of Coulomb systems and for calculation of their energy
it is natural to use the variational procedure with exponential
basis as it is most universal in respect to masses of particles
(see also \cite{FrS95}).

In general case the structure of a Coulomb system of three
single-charged particles with total charge $\pm1$ may be presented
in the form $X^\pm Y^\mp Z^\mp$ where $m_Y \le m_Z$. The stability
of the system depends on two ratios of masses, $\xi_{YX}=m_Y/m_X$
and $\xi_{ZX}=m_Z/m_X$. A boundary delimiting the regions of
stable and unstable systems is determined from the condition of
coincidence of the energy of the three-particle system with that
of the two-particle system $X^\pm Z^\mp$. The corresponding
equation determining the interdependence between $\xi_{YX}$ and
$\xi_{ZX}$ can be written as:
\begin{equation}
\label{Boundary} f(\xi_{YX},\xi_{ZX}) \equiv \frac{E(X^\pm Y^\mp
Z^\mp)}{E(X^\pm Z^\mp)}-1 = 0
\end{equation}
The solution of this equation is presented in Fig. 5 by the curve A. It is
seen that not only systems with two identical particles are stable
but also two-center systems (two heavy particles with identical
charges plus light particle with opposite charge). In contrast, a
system containing two heavy particles of opposite charges are
unstable. An exception can occur if all three particles have nearly equal
masses. This takes place for instance for exotic systems
$p^+p^-\Sigma^\pm$ ($\xi_{YX}=1$, $\xi_{ZX}=1.2749$),
$\mu^+\mu^-\pi^\pm$ ($\xi_{YZ}=1$, $\xi_{ZX}=1.3213$) and
$\pi^+\pi^-\mu^\pm$ ($\xi_{YX}=0.7568$, $\xi_{ZX}=1$) for which
$f= 0.008745$, 0.006069 and 0.002354, respectively. Of course, a
three-particle system which is stable with respect to emission of
one of the constituent particles can be unstable in the excited
state. This problem was considered in {\cite{Fro92}} for symmetric
($XYY$) systems with $m_Y/m_X \ll 1$.

For the case of systems of the type $X^{+m}Y^{+m}Z^{-m}$ containing
multiple-charged particles  the situation is quit similar to the
case of single-charged particles considered above. Among
three-body systems containing single and double charged particles
the systems of the type $X^{++}Y^-Z^+$ and $X^{++}Y^-Z^{++}$ are
unstable at any ratio of their masses, whereas the systems
$X^{++}Y^-Z^-$ are always stable. As to the systems of the type
$X^{++}Y^{--}Z^+$ they can be stable only for restricted values of
ratios of their masses. The corresponding boundary is shown in the
same Fig. 5, curve B.

\appendix
\section{Standard integrals}
A calculation of matrix elements of the Hamiltonian and its square
reduces to the evaluation of the following integrals:
\begin{equation}
\label{Iklm}
I^{klm}(x_1,x_2,x_3) \equiv 8\pi^2\int\limits_0^\infty R_1^k
dR_1\int\limits_0^\infty R_2^ldR_2\int\limits_{\mid R_1-R_2\mid}^{R_1+R_2}
R_3^mdR_3\exp \Big( -\sum\limits_{p=1}^3 x_p R_p \Big).
\end{equation}

The integrals $I^{klm}(x_1,x_2,x_3)$ with non-negative indexes are
the uniform polynomials of the $(k+l+m+3)$-th degree with respect
to the variables $A_i \equiv 1/(x_1+x_2+x_3-x_i)$.

To calculate the upper variational estimate the following
integrals are necessary:
$$
\begin{array}{rcl}
I^{000}&=&A_1A_2A_3; \\
N&=&2I^{000}\Big((A_1+A_2)(A_2+A_3)(A_3+A_1)-A_1A_2A_3\Big);
\vphantom{\Bigg\}}\\
G_1&=&I^{000}(A_1A_2+A_2A_3+A_3A_1+2A_1^2);\\
T_1&=&I^{111}-4I^{000}A_2A_3(A_2+A_3).\vphantom{\Bigg\}}
\end{array}
$$
(Here and further an unimportant numerical factor $16\pi^2$
is dropped.)

For presentation of the integrals (\ref{Iklm}) with negative indexes
it is convenient to use the following notations:
$$
\begin{array}{c}
B_1\equiv(x_2-x_3)^{-1},\hspace{1cm}
B_2\equiv(x_3-x_1)^{-1},\hspace{1cm}
B_3\equiv(x_1-x_2)^{-1};\hspace{1cm}\\

S_{C1}^{[n]}=B_1^n\ln\frac{A_2}{A_3}-B_1^{n-1}A_3^{}-\cdots-
\rel{n-1}B_1^{}A_3^{n-1};    \vphantom{\Bigg\}}    \\

S_{E1}^{[n]}=B_1^n\ln\frac{A_2}{A_3}-B_1^{n-1}A_2^{}+\cdots+
\frac{(-1)^{n-1}}{n-1}B_1^{}A_2^{n-1};        \\

S_1^{[n]}=S_{C1}^{[n]}+S_{E1}^{[n]};\vphantom{\Bigg\}}\\

N_1^{[n]}=\rel{n}\Big(A_3^n\ln\frac{A_1}{A_2}+A_2^n\ln\frac{A_1}{A_3}-
S_{C3}^{[n]}+(-1)^{n-1}S_{E2}^{[n]}\Big).
\end{array}
$$

To calculate the lower variational estimate the following integrals are necessary:
$$
\begin{array}{ccl}
J_{12}=J_{21}&=&I^{000}(A_1+A_2); \vphantom{\Bigg\}}\\
J_{11}&=&A_1^3S_1^{[1]}-A_1S_1^{[3]}+I^{000}\Big(A_1/2+(A_1+A_2+A_3)/2\Big);
\\\\

W_{12}&=&I^{000}A_1(A_3+2A_2)+2A_2^3S_2^{[2]}-
2A_2^2S_2^{[3]};\vphantom{\bigg\}}\\
W_{21}&=&I^{000}A_2(A_3+2A_1)-2A_1^3S_1^{[2]}-
2A_1^2S_1^{[3]};\vphantom{\Bigg\}}\\
W_{11}&=&G_1-2I^{000}A_2A_3;\\\\

Q_{12}=Q_{21}&=&N-4I^{000}A_3(A_1^2+A_2^2)-
8A_k^3S_3^{[3]};\vphantom{\Bigg\}}\\
Q_{11}&=&3A_2\Big( -2S_{E2}^{[1]}A_2^4 + 3S_{E2}^{[2]}A_2^3
-3S_{E2}^{[4]}A_2 + 2S_{E2}^{[5]}\Big) + \\ & + &
3A_3\Big( -2S_{C3}^{[1]}A_3^4 - 3S_{C3}^{[2]}A_3^3
\vphantom{\Bigg\}}
+3S_{C3}^{[4]}A_3 + 2S_{C3}^{[5]}\Big) + \\ & + &
(2T_1-N)+I^{5-1-1}/4;
\end{array}
$$

$$
\begin{array}{rcl}
I^{5-1-1} & = & 60\Bigg\{ x_1^{-1}\Big[N_1^{[5]}-
A_2A_3\Big(A_2^3/4+A_2^2A_3^{}/6+A_2^{}A_3^2/6+A_3^3/4\Big)\Big]+\\&+&
x_1^{-2}\Big[N_1^{[4]}-   \vphantom{\Bigg\}}
A_2A_3\Big(A_2^2/3+A_2^{}A_3^{}/4+A_3^2/3\Big)\Big]+\\&+&
x_1^{-3}\Big[N_1^{[3]}-A_2A_3\Big(A_2^{}/2+A_3^{}/2\Big)\Big]+
x_1^{-4}\Big[N_1^{[2]}-A_2A_3\Big] + \vphantom{\Bigg\}}
x_1^{-5}     N_1^{[1]} +\\&+&
x_1^{-6}\Big[ \Li(1-\frac{A_2}{A_1})+\Li(1-\frac{A_3}{A_1})
+\frac12\ln^2\frac{A_2}{A_3}+\frac{\pi^2}6 \Big]\Bigg\}.
\end{array}
$$

The expression for $I^{5-1-1}$ contains the di-logarithmic
function $\Li(z) \equiv -\int_0^z t^{-1}\ln(1-t)dt$. If
$u=x_1/x_2$ and $v=x_1/x_3$ are simultaneously small one can use
for it the expansion:
$$
I^{5-1-1} =
x_2^{-6}\sum\limits_{n=0}^{\infty}u^n P_{n+5}(w)\frac{(n+5)!}{n!(n+6)}+
x_3^{-6}\sum\limits_{n=0}^{\infty}v^n P_{n+5}(w^{-1})\frac{(n+5)!}{n!(n+6)}
$$
$$
P_q(\alpha) = P_{q-2}(\alpha) - \frac{(-\alpha)^q}{q},  \hspace{4mm}
q = 2,3,\cdots , \hspace{4mm}
P_0 = -\ln(1+\alpha) ,\hspace{4mm}
P_1 = 1, \hspace{4mm}
w = \frac{v}{u}.
$$
These formulas are used if $\max(u,v)<0.3$.

\section{Model Nuclear potentials}
Parameters of used nuclear model potentials (\ref{YukawaPot}) and (\ref{LambdaNPot}) are given
in Table 4.

In calculations of hypertritium $NN$-potentials $NN$-1 and $NN$-2
were used. The radial parameter of purely attractive potential
$NN$-1 was chosen corresponding to one-pion exchange whereas for
$NN$-2 potential the values of $R_r$ and $R_a$ were adopted  from
paper \cite{Mlf69}. The depth parameters for potentials $NN$-1 and
$NN$-2 were matched to correct deuteron energy, additional
experimental data in fitting of parameters for $NN$-2 potential
were deuteron radius and phases of S-wave triplet $np$-scattering
up to energy 300 MeV. In calculations of tritium $^3$H the
potential $NN$-3 was used with the same radial parameter as for
potential $NN$-1, whereas depth parameters was chosen to describe
the correct tritium binding energy in calculations with $N=100$ in
expansion (\ref{trial}).

The radius of $\Lambda N$-potential was adopted from paper
\cite{KTr77} while the depth parameter provided the correct
hypertritium binding energy (B$_\Lambda$ = 0.13 MeV) in
calculations with $N = 100$.


\newpage

\begin{table}

\begin{tabular}{|l|r|l|l|p{65mm}|} \hline
System        & N  & $E_U$, au        & $E_L$, au   & Comment  \\
\hline
$^\infty He$  & 10 & -2.903 723 6         & -2.903 83       &  \\
              & 30 & -2.903 724 373 0     & -2.903 725 8    &  \\
              & 50 & -2.903 724 375 9     & -2.903 725 2    &  \\
              &100 & -2.903 724 377 009   & -2.903 724 414  &  \\
              &200 & -2.903 724 377 030 3 & -2.903 724 391  &  \\
              &300 & -2.903 724 377 033 2 & -2.903 724 380  &  \\  \hline
$\alpha ee$   &300 & -2.903 304 557 732 3 & -2.903 304 561
              & $m_\alpha=7294.2996m_e$                        \\  \hline
$^\infty H^-$ & 10 & -0.527 750 546       & -0.528 062      &  \\
              & 30 & -0.527 751 009 425   & -0.527 764      &  \\
              & 50 & -0.527 751 015 895   & -0.527 752 977  &  \\
              &100 & -0.527 751 016 400   & -0.527 751 663  &  \\  \hline
$pee$         &100 & -0.527 445 880 971   & -0.527 446 533
              & $m_p=1836.1527m_e$                             \\  \hline
$\mu ee$      &100 & -0.525 054 806 098   & -0.525 055 501
              & $m_\mu=206.768262m_e$                          \\  \hline
$Ps^-$        & 10 & -0.262 003 563       & -0.262 744      &  \\
              & 30 & -0.262 005 053       & -0.262 026      &  \\
              & 50 & -0.262 005 068 6     & -0.262 008 7    &  \\  \hline
$pp\mu$       & 10 & -0.494 374           & -0.495 7
              & In meso-atomic units                           \\
              & 30 & -0.494 386 645       & -0.494 408
              & $m_p=8.8802444m_\mu$                           \\
              & 50 & -0.494 386 790       & -0.494 391 1    &  \\  \hline
$dd\mu$       & 10 & -0.531 044           & -0.534 4
              & In meso-atomic units                           \\
              & 30 & -0.531 109 463       & -0.531 241
              & $m_d=17.7516751m_\mu$                               \\  \hline
$tt\mu$       & 10 & -0.546 224           & -0.551 2
              & In meso-atomic units                           \\
              & 30 & -0.546 371 871       & -0.546 517
              & $m_t=26.5849388m_\mu$                               \\  \hline
$\mu\mu e$    & 10 & -0.583 276           & -0.604 5        &  \\
              & 30 & -0.584 757           & -0.588 82       &  \\
              & 50 & -0.584 995           & -0.586 267      &  \\
              & 20 & -0.585 126 081 8     &
              & 'Carcass' basis                                \\  \hline
$ppe$         & 10 & -0.591 03            & -0.625          &  \\
              & 30 & -0.595 02            & -0.612          &  \\
              & 50 & -0.595 67            & -0.606          &  \\
              & 20 & -0.597 139 058 5     &
              & 'Carcass' basis                                \\  \hline
$dde$         & 10 & -0.591 38            & -0.621          &  \\
              & 30 & -0.596 06            & -0.610          &  \\
              & 20 & -0.598 788 780 3     &
              & 'Carcass' basis                                \\  \hline
$tte$         & 10 & -0.591 59            & -0.615          &  \\
              & 30 & -0.596 34            & -0.608          &  \\
              & 20 & -0.599 506 906 3     &
              & 'Carcass' basis                                \\  \hline
$p\mu e$      & 10 & -0.584 18            & -0.645          &
\\  \hline $\alpha\mu e$ & 10 & -1.947 287 542       & -1.947 429
              & In meso-atomic units                           \\
              & 30 & -1.947 287 553 22    & -1.947 290 320
              & $E_0=$\mbox{-1.947~287~553~40},

                $m_\alpha=35.2776559m_\mu$,

                $m_e=0.00483633218m_\mu$                       \\  \hline
$^3_\Lambda$H
              & 50 & -2.359 478 5         & -2.437 6
              & Potentials NN-1 and $\Lambda$N-1; energies are in MeV  \\  \hline
$^3_\Lambda$H
              & 50 & -2.358 597 8         & -2.808
              & Potentials NN-2 and $\Lambda$N-2; energies are in MeV \\  \hline
$^3$H         & 50 & -8.480 037 312       & -8.480 045 6
              & Potential NN-3; energies are in MeV                   \\  \hline
\end{tabular}
\caption{Upper and lower bounds for Coulomb and nuclear systems}
\end{table}
\newpage

\begin{table}
\begin{tabular}{|l|c|c|c|c|} \hline
Variant      & $V_r$, MeV & $R_r$, Fm & $V_a$, MeV & $R_a$, Fm \\
\hline
NN-1         & 0          &    0      & 50.6414    & 1.4       \\
NN-2         & 2719.20    & 0.32      & 730.24     & 0.65      \\
NN-3         & 0          &    0      & 40.0419    & 1.4       \\
$\Lambda$N-1 & 0          &    0      & 687.00     & 0.23      \\
$\Lambda$N-2 & 0          &    0      & 711.00     & 0.23      \\
\hline
\end{tabular}
\caption{ Parameters of nuclear model potentials}
\end{table}

\newpage
\begin{figure}
\centering
\includegraphics[width=10cm]{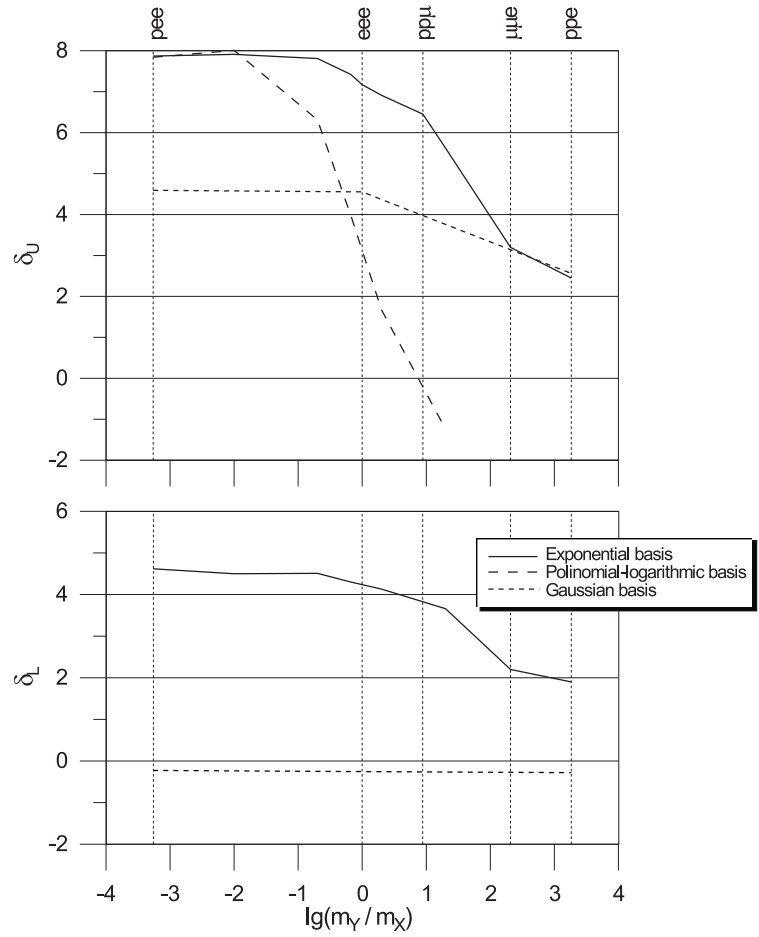}
\caption{ Dependence of $\delta_U$ and $\delta_L$ on mass ratio
for Coulomb systems $X^+Y^-Y^-$}
\end{figure}

\begin{figure}
\centering
\includegraphics[width=10cm]{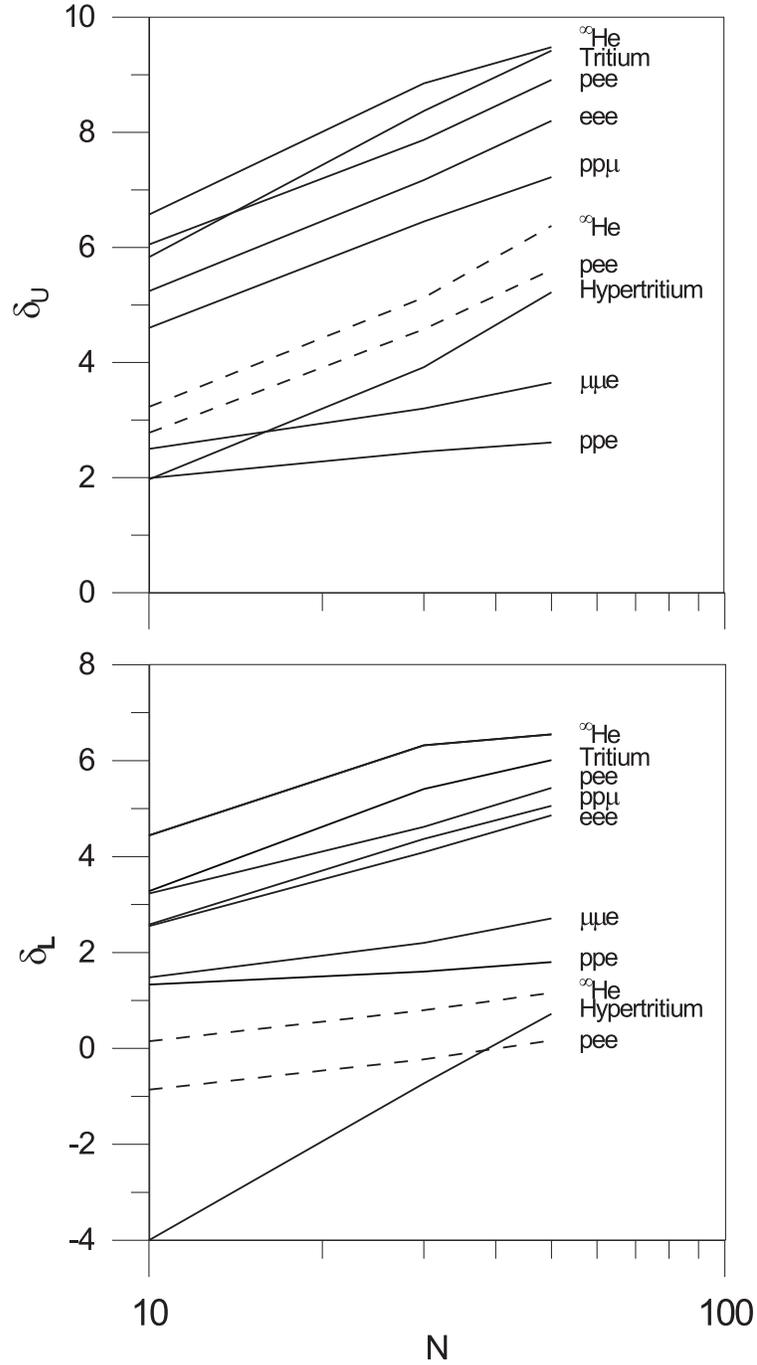}
\caption{Dependence of $\delta_U$ and $\delta_L$ on number of
terms in variational expansion for exponential basis}
\end{figure}

\begin{figure}
\centering
\includegraphics[width=10cm]{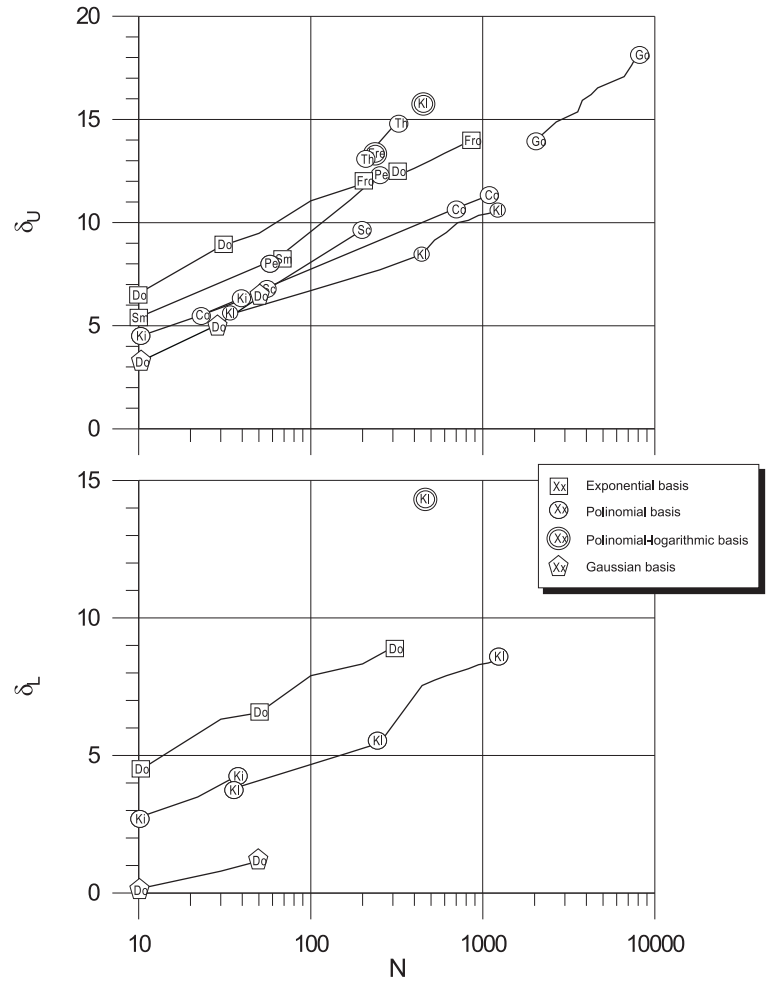}
\caption{$\delta_U$ and $\delta_L$ in calculations of atom
$^\infty$He with different variational expansions. Markers refer
to the first author of the corresponding paper.}
\end{figure}

\begin{figure}
\centering
\includegraphics[width=10cm]{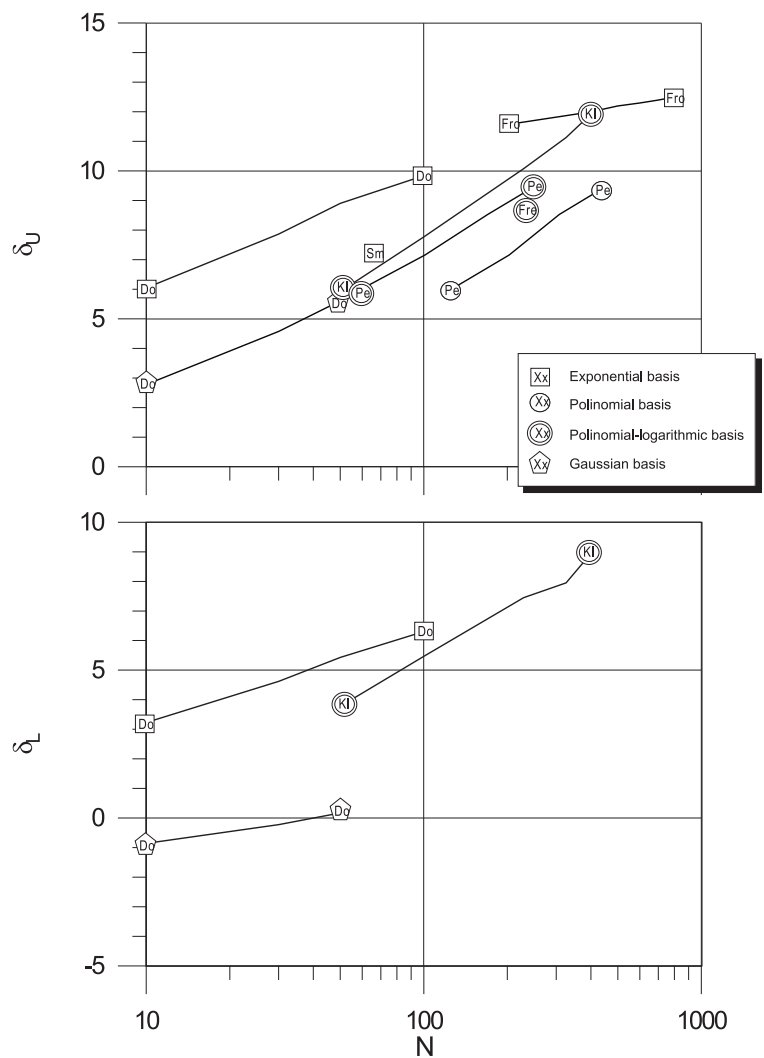}
\caption{ The same as in Fig. 3 but for hydrogen ion}
$^\infty$H$^-$
\end{figure}

\begin{figure}
\centering
\includegraphics[width=10cm]{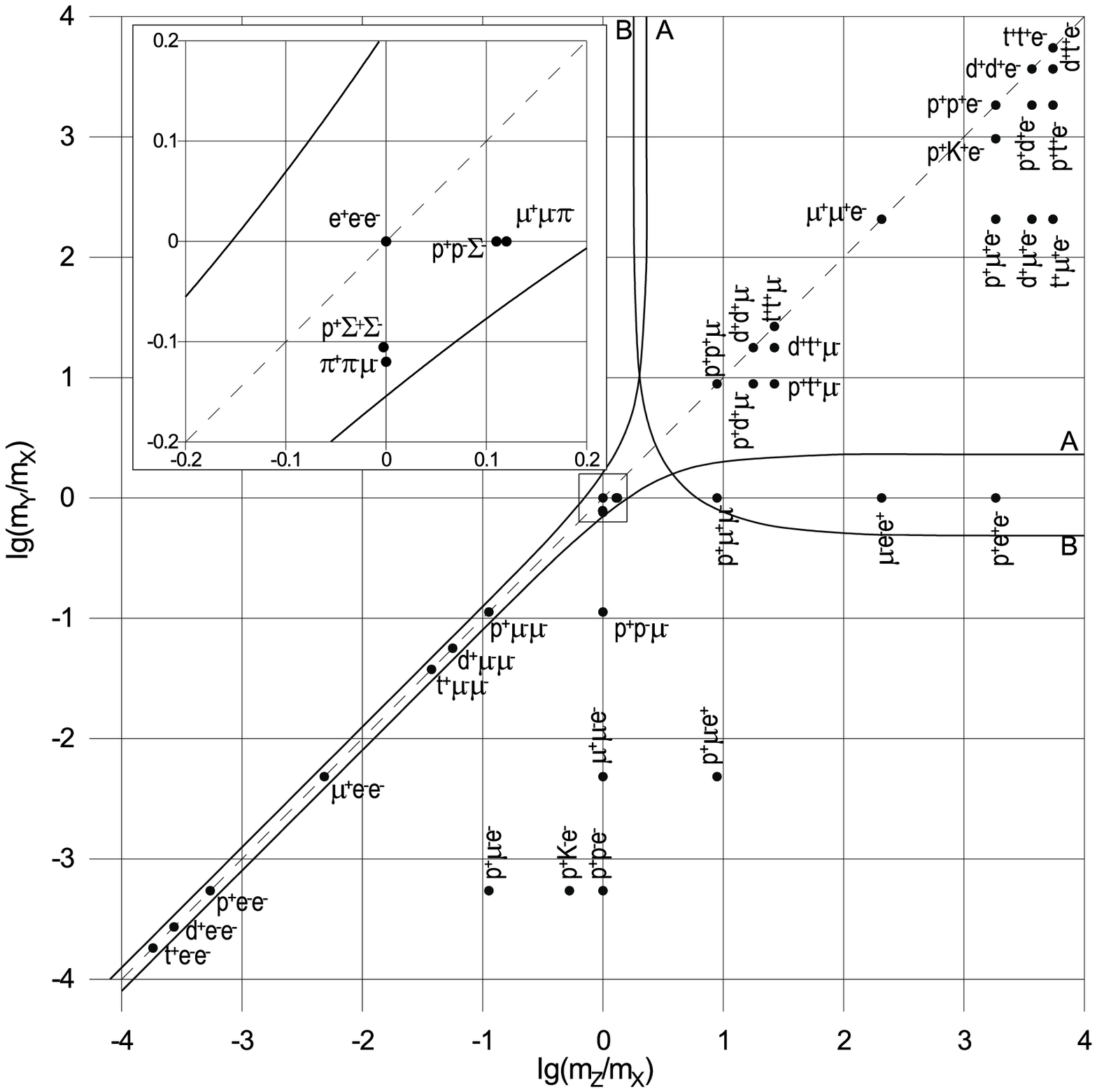}
\caption{ Boundary of stability for 3-particle Coulomb systems.}
\end{figure}


\begin{thebibliography}{99}

\bibitem{Sch62}
Schwartz C.: Phys.Rev. {\bf 128}, 1147 (1962)

\bibitem{Thk94}
Thakkar A.J., Koga T.: Phys.Rev. {\bf A50}, 854 (1994)

\bibitem{Pkr66}
Frankowski K., Pekeris C.L.: Phys.Rev. {\bf 146}, 46 (1966)

\bibitem{Fro92}
Frolov A.M.: J.Phys. B: At.Mol.Opt.Phys. {\bf 25}, 3059 (1992)

\bibitem{KTr82}
Kolesnikov N.N., Tarasov V.I.: J.Nucl.Phys. {\bf 35}, 609 (1982)

\bibitem{Vrg98}
Usukura J., Varga K., Suzuki Y.: Phys.Rev. {\bf A58}, 1918 (1998)

\bibitem{Don00}
Donchev A.G., Kolesnikov N.N., Tarasov V.I.: Phys. At. Nucl. {\bf
63}, 419 (2000)

\bibitem{Kln90}
Kleindienst H., Emrich R.: Int.J.Quant.Chem. {\bf 37}, 257 (1990)

\bibitem{Kin57}
Kinoshita T.: Phys.Rev. {\bf 105}, 1490 (1957)

\bibitem{Pkr62}
Pekeris C.L.: Phys.Rev. {\bf 126}, 1470 (1962)

\bibitem{Smt77}
Thakkar A.J., Smith V.H., Jr: Phys.Rev. {\bf A15}, 1 (1977)

\bibitem{Kln80}
Kleindienst H., Wolfgang M.: Theoret.Chim.Acta. {\bf 56}, 183
(1980)

\bibitem{Fnd84}
Freund D.E., Huxtable B.D., Morgan III J.D.: Phys.Rev. {\bf A29},
980 (1984)

\bibitem{Cox94}
Cox H., Smith S.J., Sutcliffe B.T.: Phys.Rev. {\bf 49}, 4520
(1994); ibid {\bf 49}, 4533 (1994)

\bibitem{Gld98}
Goldman S.P.: Phys.Rev. {\bf A57}, 677 (1998)

\bibitem{Frl98}
Frolov A.M.: Phys.Rev. {\bf A58}, 4479 (1998)

\bibitem{Kms95}
Komasa J., Cencek W., Pychlewski J.: Phys.Rev. {\bf A52}, 4500
(1995)

\bibitem{Frl97}
Frolov A.M., Smith V.H., Jr.: Phys.Rev. {\bf A55}, 2662 (1997)

\bibitem{Yan98}
Yan Z.-C., Tambasco M., Drake G.W.: Phys.Rev. {\bf A57}, 1652
(1998)

\bibitem{Yan99}
Yan Z.-C., Ho Y.K.: Phys.Rev. {\bf A59}, 2697 (1999)

\bibitem{Frl99}
Frolov A.M.: Phys.Rev. {\bf  A60}, 2834 (1999)

\bibitem{FrS98}
Frolov A.M.: Phys.Rev. {\bf A57}, 2436 (1998); ibid {\bf A59},
4270 (1999)

\bibitem{BOp27}
Born M., Oppenheimer J.R.: Ann.Phys. {\bf 84}, 457 (1927)

\bibitem{Pon81}
Ponomarev L.I.: J.Phys. {\bf B14}, 591 (1981)

\bibitem{GDB98}
Gremaud B., Dominique D., Billy N.: J.Phys. {\bf B31}, 383 (1998)

\bibitem{Zah83}
Zakharov P.P., Kolesnikov N.N., Tarasov V.I.: Vestn.Mosk.Univ.
Ser.3: Fiz,Astron. {\bf 24}, 34 (1983), in russian

\bibitem{FrS95}
Frolov A.M., Smith V.H., Jr. V.H.: J.Phys. B: At.Mol.Opt.Phys.
{\bf 28}, L449 (1995)

\bibitem{Mlf69}
Malfliet R.A., Tjon J.A.: Nucl.Phys. {\bf A127}, 161 (1969)

\bibitem{KTr77}
Kolesnikov N.N., Tarasov V.I.: Vestn.Mosk.Univ. Ser.3: Fiz,Astron.
{\bf 18}, 8 (1977), in russian


\end{thebibliography}
\end{document}